\documentclass[aps, prb, twocolumn, oneside, 10pt, amsfonts, amsmath, nofootinbib, floatfix, superscriptaddress]{revtex4-2}
\usepackage{graphicx}
\usepackage[colorlinks=true, citecolor = red]{hyperref}
\usepackage{physics}
\usepackage{hhline}
\usepackage{caption}
\usepackage{subcaption}
\usepackage{footnote}
\usepackage{blkarray}
\usepackage{kbordermatrix} % include package @ document preamble
\newcommand{\av}{{\bf a}}
\newcommand{\kv}{{\bf k}}

\renewcommand{\kbldelim}{(}
\renewcommand{\kbrdelim}{)}
\usepackage{float}
\usepackage{xcolor}
\usepackage{wasysym}
\newcommand{\Hzz}{$H_{00}$} 
\newcommand{\Hhh}{$H_{\frac{1}{2}\frac{1}{2}}$} 
\newcommand{\SEHzz}{$SEH_{00}$} 
\newcommand{\SEHhh}{$SEH_{\frac{1}{2}\frac{1}{2}}$} 

\begin{document}
\title{Designing aperiodic to periodic interfaces}
\author{Sam Coates}
\affiliation{Surface Science Research Centre and Department of Physics, University of Liverpool, Liverpool L69 3BX, UK}
\email{Corresponding author: samuel.coates@liverpool.ac.uk}

\begin{abstract}
	
Symmetry sharing facilitates coherent interfaces which can transition from periodic to aperiodic structures. Motivated by the design and construction of such systems, we present hexagonal aperiodic tilings with a single edge-length which can be considered as decorations of a periodic lattice. We introduce these tilings by modifying an existing family of golden-mean trigonal and hexagonal tilings, and discuss their properties in terms of this wider family. Then, we show how the vertices of these new systems can be considered as decorations or sublattice sets of a periodic triangular lattice, before introducing methods to designing coherent aperiodic to periodic interfaces. 

%We conclude by simulating a simple Ising model on one of these decorations, and compare this system to a triangular lattice with random defects.

\end{abstract}
\date{\today}
\maketitle
 
\section{Introduction and motivation}

Aperiodic tilings have been well-studied across the range of the physical sciences. They can be used as simple tools to better understand or generate quasicrystalline phases of matter, studied as intrinsic mathematical objects, or utilised in applied research as the basis for fabricated/manipulated structures. Focus has been applied to such tilings which exhibit rotational symmetries which are incommensurate with periodicity or translational symmetry: 5-, or greater than 6-fold. However, this is not a required condition -- aperiodic tilings can, of course, be 2-, 3-, 4-, or 6-fold \cite{Robinson1971undecidability,Berger1966undecidability,Sasisekharan1989non,Clark1991quasiperiodic,Lifshitz2002square,Lifshitz2003quasicrystals,Lifshitz2007crystal,Socolar11,Dotera17,Coates2024hexagonal,Smith2023aperiodic}. 

The sharing of symmetries opens the door to coherent periodic-to-aperiodic interfaces, as it is the gateway to reducing or minimizing structural frustration. For example: decorating a 3-fold symmetric lattice with a 5-fold symmetric pattern, or sandwiching together 3-fold and 5-fold structures results in heterogeneous or incommensurate interactions between the two systems. On the other hand, periodic and aperiodic structures could mix in a more systematic manner if their local environments are cohesive, which represents a novel research direction. The motivation for this work stems from this idea, where the ultimate goal is to demonstrate routes for the design of a minimally frustrated aperiodic-periodic interface. Here, we focus on using 6-fold aperiodic tilings as a base, which we choose as hexagonal (and trigonal) structures have the highest coordination number for a 2D periodic lattice (6), and therefore offer a more `flexible' local neighbourhood for manipulation.

Previously \cite{Coates2024hexagonal}, we have introduced a family of hexagonal and trigonal tilings using a generalised version of de Bruijn's dual grid method \cite{deBruijn81, deBruijn86, Socolar85, Gahler86, Ho86, Rabson88, Rabson89, Lifshitz05, Coates2024hexagonal}. The structure of the tilings could be controlled by two parameters, $\alpha_s$ and $\alpha_l$, such that we labelled our tilings as $H_{\alpha_s \alpha_l}$. Each tiling consisted of hexagonal and/or rhombic tiles with edge lengths built by the linear combination of two length-scales: 1 and $\tau = \frac{1+\sqrt{5}}{2}$. In this work we paid particular attention to two `special' cases in this family, which we referred to as \Hzz{} and \Hhh{}. 

Here, we modify the construction of the \Hzz{} and \Hhh{} systems with a view to generate single edge-length hexagonal ($SEH$) aperiodic tilings. The single edge-length parameter more readily leads to periodic-to-aperiodic interfaces with relatively simple local environments -- and is an advantage for both physical and theoretical realisations. First, we show the basic method used to construct the $SEH$ tilings and compare them to their \Hzz{} and \Hhh{} counterparts. Then, we present the structural properties of the \SEHzz{} and \SEHhh{} tilings in their own right in terms of their vertices and discuss their relationship to periodic lattices. Next, we demonstrate how to design and build aperiodic to periodic interfaces using two different methods, and briefly define a new aperiodic tiling. We conclude by discussing the many variations that can be constructed by deliberate and/or creative choices, and identify potential applications for our systems across multiple length-scales and disciplines.

\section{Construction and properties of the\textit{ SEH} tilings \label{sec:SEH}}

Our previous work presented a comprehensive description of the dual grid method we used to create the $H_{\alpha_s \alpha_l}$ tilings \cite{Coates2024hexagonal}. Here, we very briefly discuss the general concept, introduce the relevant parameters for the formation of the $SEH$ tilings, and discuss some properties. 

\subsection*{Dual grid method and the \textbf{$H$} tilings}

\begin{figure*}
	\centering
	\includegraphics[width=\linewidth]{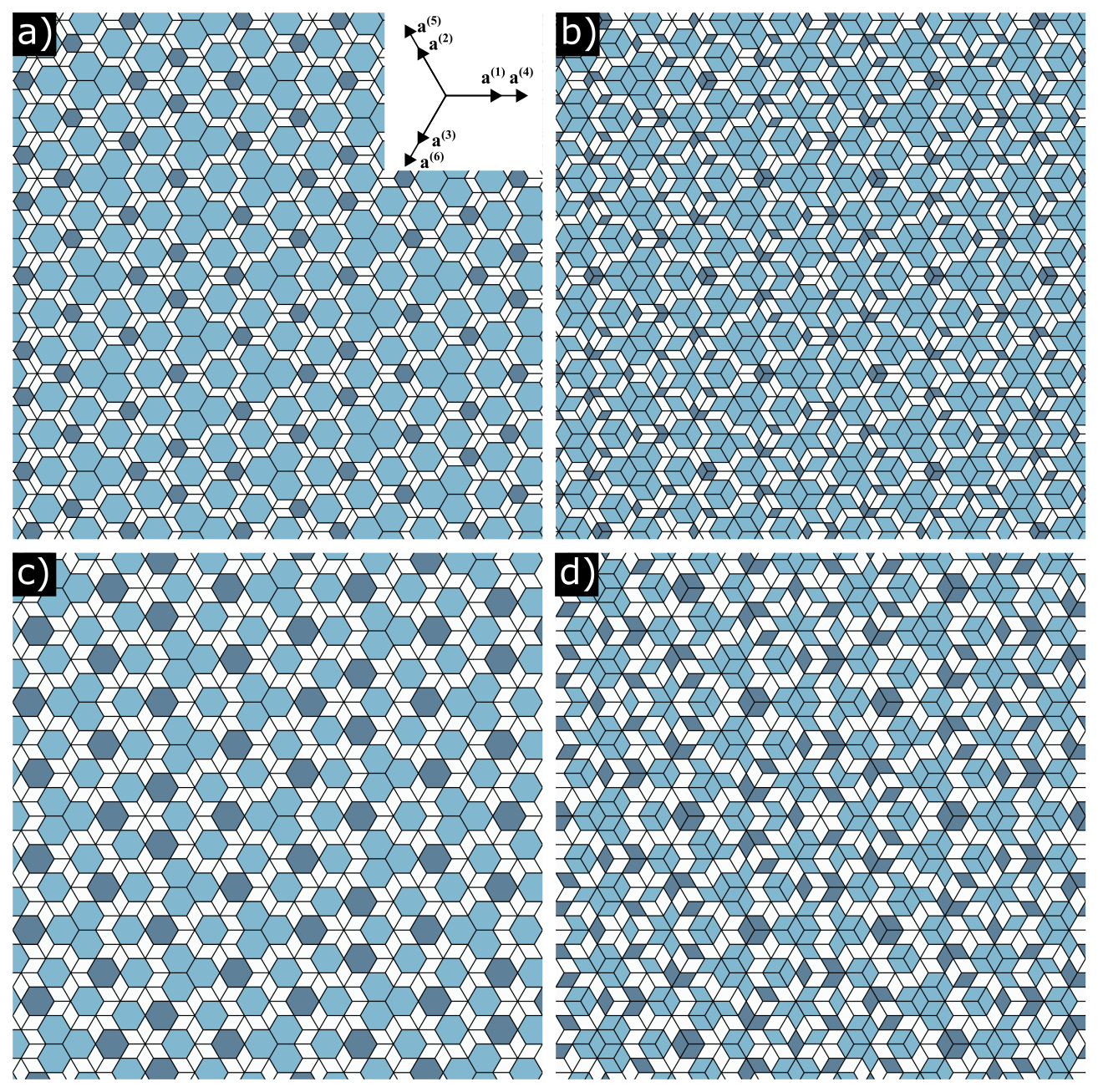}
	\caption{\textbf{(a)} The \Hzz{} tiling, where $\av^{(1-3)}$ are a factor of $\tau$ shorter than $\av^{(4-6)}$, and the shifts applied to the grids sum to 0. \textbf{(b)} The \Hhh{} tiling, where the shifts applied to the grids sum to 0.5. \textbf{(c)} The \SEHzz{} tiling, where the length of all $\av = 1$, and the shifts applied to the grids sum to 0. \textbf{(d)} The \SEHhh{} tiling, where the shifts applied to the grids sum to 0.5.\label{fig:tilings}}
\end{figure*}
An infinite set of regularly-spaced parallel lines defines a grid, where the spacing and orientation of the grid lines is determined by a perpendicular grid vector $\kv^{(j)}$. A multigrid is then composed of a set of superimposed grids, with grid vectors $\kv$. In the simplest terms, the placement or arrangement of tiles in physical space is determined by the intersection points of this multigrid. The distribution of intersection points can be changed via fractional translational shifts applied to the grids, $f_j$, along the direction of the grid vectors $\kv^{(j)}$. We parameterised our family of tilings considering the sum of our grid shift components, such that 

\begin{equation}\label{Eq:invariants}
	\alpha_s=f_1+f_2+f_3\quad \textrm{and}\quad \alpha_l=f_4+f_5+f_6,
\end{equation}

\noindent which lead us to the $H_{\alpha_s \alpha_l}$ nomenclature, and the \Hzz{} and \Hhh{} tilings as discussed \cite{Coates2024hexagonal}. Finally, in addition to $\kv$, we have an associated set of tiling vectors, $\av$, which build our constituent tiles once we find our intersection points. More in-depth discussions on the dual grid method can be found elsewhere \cite{deBruijn81, deBruijn86, Socolar85, Gahler86, Ho86, Rabson88, Rabson89, Lifshitz05, Coates2024hexagonal}.

To create the \Hzz{} and \Hhh{} tilings, we separated our grid and tiling vectors into two groups, choosing $\av^{(1-3)}$ to be a factor of $\tau$ shorter than $\av^{(4-6)}$ (inset of Figure \ref{fig:tilings}), and the spacing between grids $\kv^{(1-3)}$ to be a factor of $\tau$ longer than $\kv^{(4-6)}$. Figures \ref{fig:tilings}(a) and (b) show examples of the resultant tilings, respectively. The \Hzz{} tiling is comprised of 3 tiles: a small hexagon (edge length = 1), a parallelogram (edge lengths = 1, $\tau$), and a large hexagon (edge length = $\tau$). The \Hhh{} tiling is technically built using two mirror-symmetric parallelogram tiles, three small, and three large rhombuses. However, the colour scheme of Figure \ref{fig:tilings}(b) is simplified as the specific properties of these tiles are not discussed here. 

The arrangement of tiles in the \Hzz{} and \Hhh{} tilings is aperiodic due to the irrational scaling factor that we set between two families of grid vectors $\kv^{(1-3)}$ and $\kv^{(4-6)}$. Trivially, the irrational scaling guarantees no periodicity in the set of intersection points we create. The tiling vectors, however, only affect the geometry of the tiles, and as such their scale can be chosen freely \cite{Gahler86}. In our original work, we chose to reflect the $\tau$ scaling in our choice of tiling vectors. However, here, we can simply set the scaling factor of the tiling vectors to be 1 and produce aperiodic arrangements of single edge-length tiles. 
\subsection*{The \textit{SEH} tilings}

Keeping our focus on systems where $\alpha_s \equiv \alpha_l \equiv 0$, and $\alpha_s \equiv \alpha_l \equiv 0.5$, Figures \ref{fig:tilings}(c, d) show the \SEHzz{} and \SEHhh{} tilings, which are generated by keeping the scaling factor between $\kv^{(1-3)}$ and $\kv^{(4-6)}$ as $\tau$, and setting the lengths of $\av = 1$. The colour scheme of individual tiles is kept constant under this change, for clarity. Decreasing the scale of tiling vectors $\av^{(4-6)}$ decreases the longer parallelogram edges of \Hzz{} and \Hhh{} to form white rhombuses, while the larger hexagons in \Hzz{} and rhombuses in \Hhh{} shrink to match the size of their smaller counterparts.

Comparing the $H$ and $SEH$ tilings, it is trivial that all scale-independent properties such as tile frequency, edge-matching rules, vertex frequencies etc. still hold. However, the substitution rules we have previously discussed for the $H$ tilings are no longer valid \cite{Coates2024hexagonal}, as we have lost the self-similar scaling factor of $\tau$; whether the SEH tilings have direct substitution rules is an open topic. However, of course, substitution can be achieved via a two-step process by elongating the appropriate edges of tiles constructed by $\av^{(4-6)}$ by $\tau$, applying the substitution rules derived in \cite{Coates2024hexagonal}, and then re-scaling $\av^{(4-6)}$ back to 1. On a related note, isolated hexagonal tiles i.e., those not connected to other hexagons by an edge, can be decorated with a \Hzz{} tiling where the ratio of the long to short edge lengths of tiles is 3/2, an approximation of $\tau$. When we decorate the same `parental' tiles in the $\tau$-scale \Hzz{} tiling, we find it forms a $\tau^2$ inflated version of the original, where the tiles have edge lengths 2+$\tau$ and 1+$\tau$. Therefore, as we shrink $\tau \rightarrow 1$ to create the \SEHzz{} tiling, we create a 3:2 ratio. A similar mapping likely holds for the \SEHhh{} tiling.

%\section{\textit{SEH} ti/ling vertices and periodic lattice decorations}
\begin{figure*}
	\centering
	\includegraphics[width=.9\linewidth]{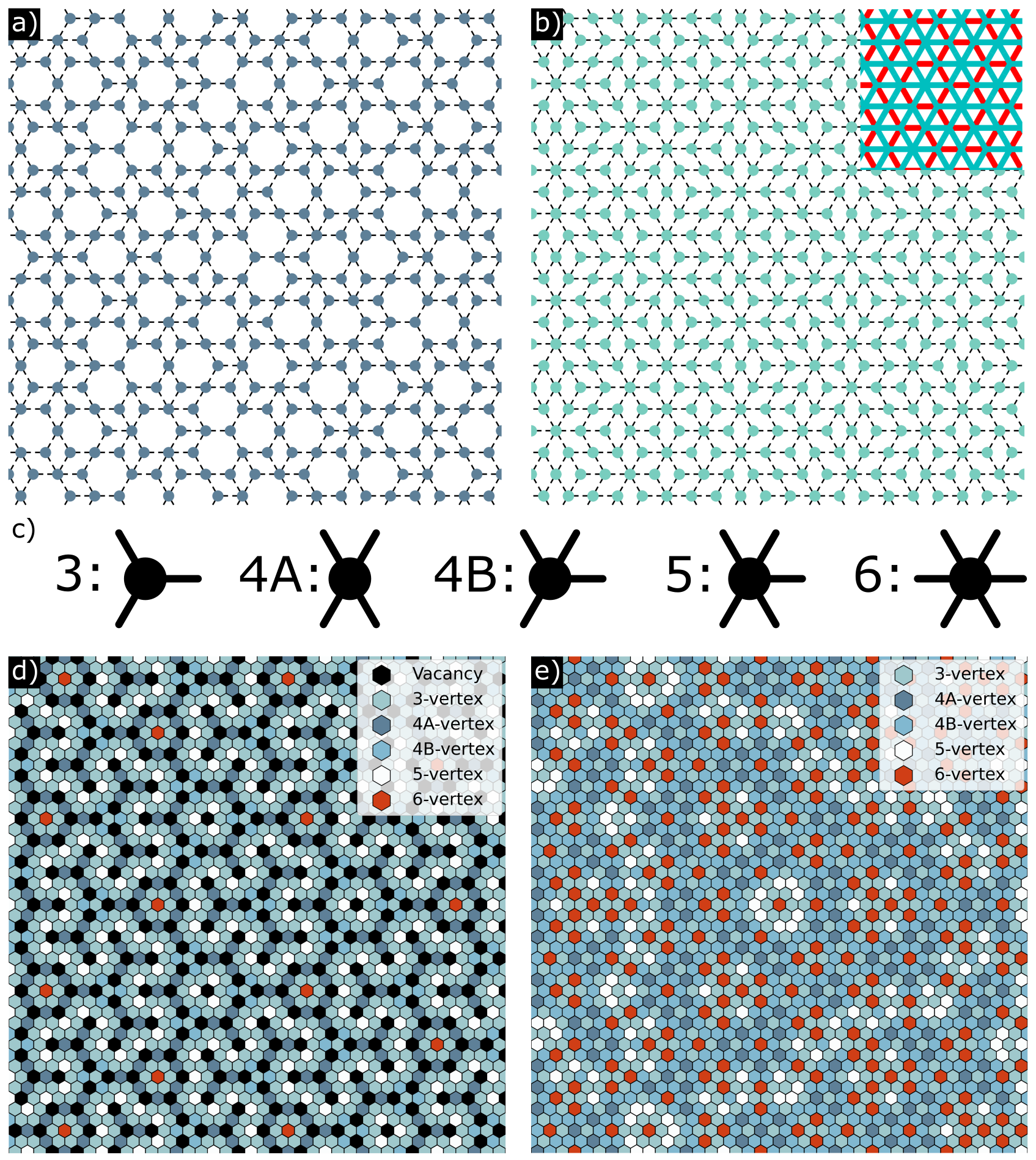}
	\caption{\textbf{(a, b)} The vertex schematics of the \SEHzz{} and \SEHhh tilings, where the edges of tiles have been bisected. Under this scheme, the tilings can be considered as positions by `bonds'. Inset in \textbf{(b)} is an overlay of a periodic triangular tiling (cyan), where certain edges are coloured red. These edges, when removed, form the \SEHhh{} tiling. \textbf{(c)} The five $n$-vertex types associated with the tilings under the bond picture. Each type is labelled according to their coordination number. \textbf{(d, e)} $n$-vertex models of the \SEHzz{} and \SEHhh{} tilings, respectively. Here, each vertex has been colour-coded depending on their coordination number, with an additional vacancy position added at the centre of the hexagonal tiles in the \SEHzz{} tiling. The points occupy separate sublattices of a periodic triangular lattice.\label{fig:skele}}
\end{figure*}

\subsection*{Vertex properties}
%The top of Figure \ref{fig:overlay}(b) shows the seven vertex configurations of the \Hzz{} tiling \cite{Coates2024hexagonal}, and the bottom shows the corresponding vertices in the \SEHzz{} tiling.
% -- changing the tile edge lengths does not affect the number of vertex types, nor their frequency across the tiling, as these are determined in grid-space. The same is true for the \Hhh{} tiling -- for conciseness however we do not show these; the \Hhh{} tiling has 32 vertex types.  T

Analysis of vertices in a aperiodic tiling is commonly done by considering the points as projections from a higher-dimensional superspace. Indeed, we previously showed how both the \Hzz{} and \Hhh{} tilings can be constructed via the projection of a hypercubic lattice. The basis of this lattice was defined by a matrix of six orthogonal 6-dimensional vectors, whose first two rows contain the tiling vectors \av{} [\cite{Coates2024hexagonal}, section IV]. For the $SEH$ tilings, we can still view the vertices as projections onto an internal subspace using the same matrix, such that the internal subspace windows and their subdivisions are identical to the $H$ tilings. In this case, we do not obtain any new information by considering the $SEH$ tilings in hyperspace. However, as we have altered \av, the original matrix no longer describes an orthogonal hypercubic lattice.

%\begin{figure*}
%	\centering
%	\includegraphics[width=\linewidth]{images/verts}
%	\caption{\textbf{(a, b)} $n$-vertex models of the \SEHzz{} and \SEHhh{} tilings, respectively. Here, each vertex has been colour-coded depending on their coordination number, with an additional vacancy position added at the centre of the hexagonal tiles in the \SEHzz{} tiling. The points occupy separate sublattices of a periodic triangular lattice. \label{fig:verts}}
%\end{figure*}

\subsection*{\textit{SEH} vertices as periodic lattice decorations}

Figures \ref{fig:skele}(a, b) show the vertices of the \SEHzz{} and \SEHhh{} tilings respectively, where the edges of the tiles have been bisected to illustrate the relationship to periodic lattices and the connectivity between adjacent vertices. This relationship is clearest for the \SEHhh{} tiling, where each vertex perfectly decorates a periodic triangular lattice with a lattice constant of 1. Under this scheme, it becomes clear that the tile edges or bonds between vertices are aperiodic, while the vertex or point distribution is not. Therefore, an alternative way to produce the \SEHhh{} tiling would be to remove a subset of edges from a periodic triangular lattice. An example is overlaid and inset in Figure \ref{fig:skele}(b), where the cyan lines indicate the triangular lattice, and the red lines are edges which are removed. We discuss the finer points of this method in Appendix \ref{app:deletion}. 

\begin{figure*}
	\centering
	\includegraphics[width=\linewidth]{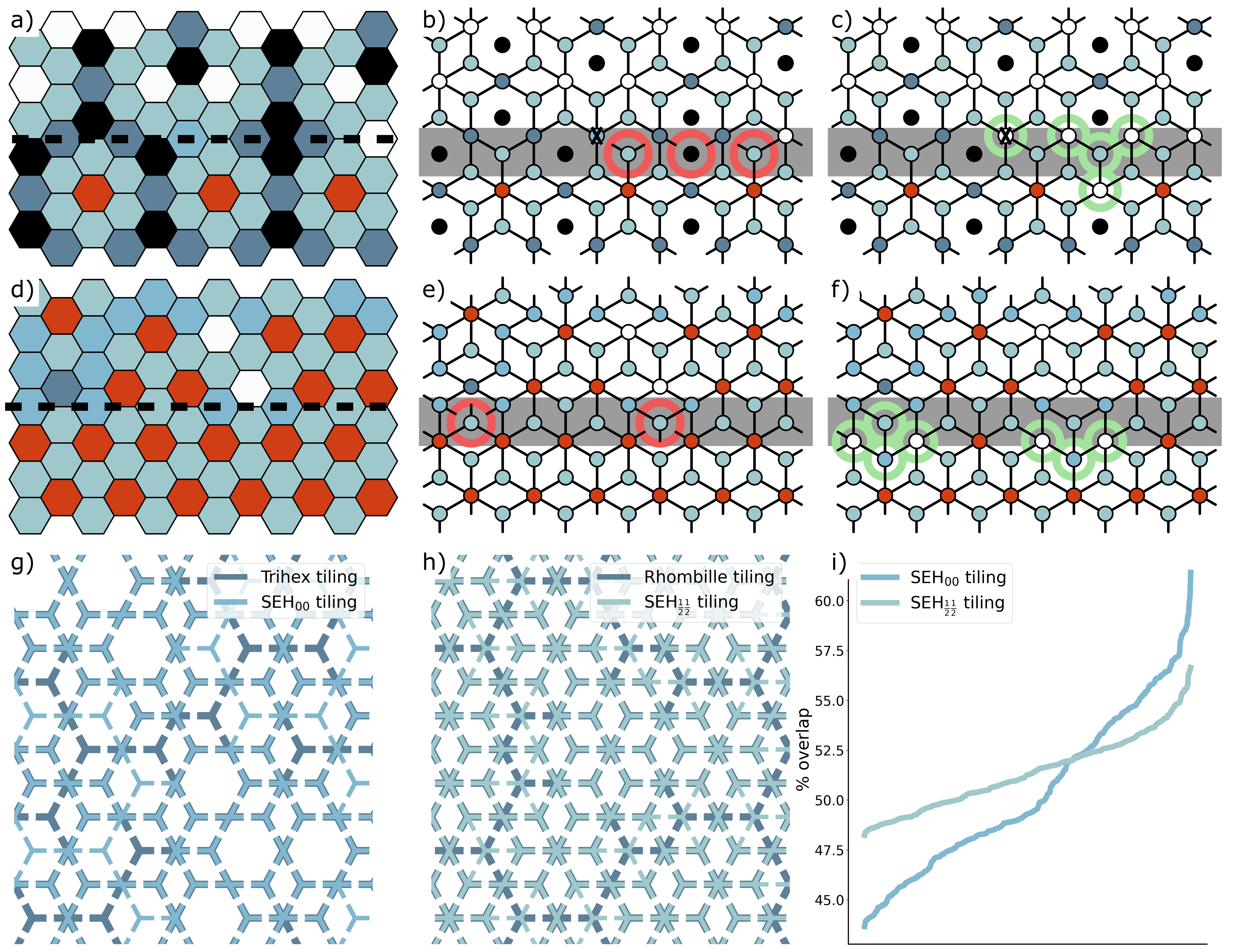}
	\caption{\textbf{(a)} Vertex maps for a section of the \SEHzz{} tiling (top) and its periodic analogue (bottom), where a dashed line indicates the interface formed between the two. \textbf{(b)} Bond map of \textbf{(a)}, where broken bonds are highlighted with red circles, and the interface is marked by a grey bar. A 4B-vertex is indicated with an `X'. \textbf{(c)} The broken bonds of \textbf{(b)} are mended by changing vertex types, highlighted by green circles. The 4B-vertex marked in \textbf{(b)} has now changed to a 5-vertex, for example. \textbf{(d--f)} Same as \textbf{(a--c)}, but for the \SEHhh{} system. \textbf{(g, h)} Patches of the bond maps of the \SEHzz{} and \SEHhh{} tilings with maximum overlap with their periodic analogues. \textbf{(i)} Percentage overlap of the two systems with their periodic analogues as the centre of an aperiodic disc of radius 25 is translated atop a periodic disc of radius 50. \label{fig:interpolate}}
\end{figure*}

\begin{figure*}
	\centering
	\includegraphics[width=\linewidth]{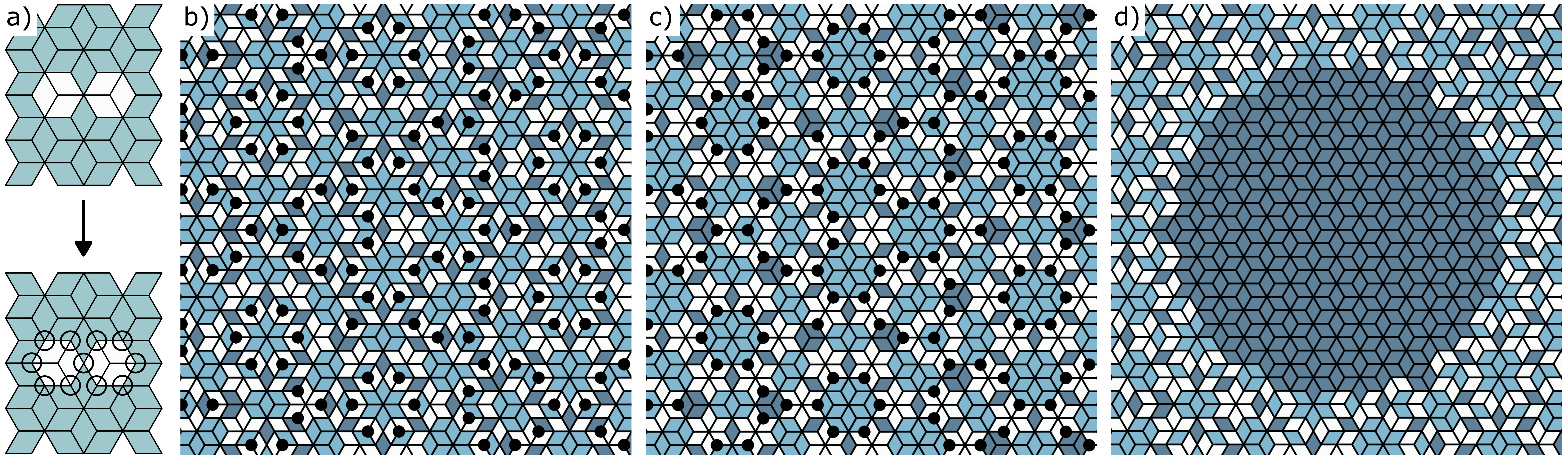}
	\caption{\textbf{(a)} Top: a section of the periodic rhombille tiling, with two 3-vertices indicated by the white-shaded tiles. Bottom: after rotating these vertices by 180$^\circ$, new vertex configurations are created, as indicated by the black circles. \textbf{(b)} A section of the \Hhh{} tiling, with a subset of 3-vertices highlighted by black circles. \textbf{(c)} The same section of the \Hhh{} tiling as in \textbf{(b)}, where the 3-vertices have been rotated by 180$^\circ$. \textbf{(d)} After rotating the subset of 3-vertices of a \Hhh{} tiling within a radius of 10 from the origin, we form an interface with the periodic rhombille tiling. \label{fig:rhombille}}
\end{figure*}

After bisecting our tile edges, we can classify the vertices of the $SEH$ tilings with respect to their coordination number. The five vertex types are shown in Figure \ref{fig:skele}(c); we refer to them as $n$-vertices, where $n$ is their coordination number. We label the two 4-vertices as 4A and 4B, as they are geometrically unique. Through this scheme, we can consider each set of $n$-vertices as an aperiodic subset which occupies a periodic triangular lattice. Figures \ref{fig:skele}(d, e) show the \SEHzz{} and \SEHhh{} distributions of the  $n$-vertices respectively, colour-coded with respect to their $n$-vertex type, and plotted as hexagons for clarity. For the \SEHzz{} tiling, we plot the centre of the hexagonal tiles as vacancies.

\section{Building an aperiodic to periodic interface}

Here, we discuss two different methods for creating an aperiodic to periodic interface, and present a new tiling inspired by one of these methods.
%For reference, we plot the radial distribution functions for each vertex type across the two tilings in Appendix \ref{app:RDFs}.
\subsection*{Sandwiching vertex maps}
From a design perspective, the vertex maps in Figures \ref{fig:skele}(d, e) represent the first step in creating aperiodic to periodic interfaces. As they decorate a triangular lattice, all we need are analogous vertex maps of periodic systems: from here, we can simply sandwich segments of the vertex maps together. The incommensurate nature of the bond distributions between the two segments means that adjacent vertices will not always connect, creating defects. However, as a second step, we can replace these vertices with one of the other 5 types. As such, we can create a seamless interface.

Figure \ref{fig:interpolate}(a) shows two segments of vertex maps: above the dashed line is a segment from the \SEHzz{} tiling, below the line is a map segment from a periodic equivalent: a trihexagonal tiling consisting of hexagons and 6 sets of rhombi arranged in a star. These maps are rotated 30$^\circ$ relative to Figure \ref{fig:skele}(d,e) for clarity. Figure \ref{fig:interpolate}(b) shows the bond connectivity of this sandwich - each vertex is coloured as in Figure \ref{fig:interpolate}(a). The interface is indicated by the grey bar, and the resultant defects caused by mismatching bonds are circled in red. Figure \ref{fig:interpolate}(c) shows one `solution' to removing or repairing these bonds, where several nearby vertices have been switched from one type to another, indicated by the green circles. Figures \ref{fig:interpolate}(d--f) show an analogous system for the \SEHhh{} tiling, where its periodic counterpart is the rhombille or `dice' tiling. Neither solution is unique, and are only presented to demonstrate the bond repairing process: in both cases, a seamless interface is created between the two structures.

Changing a single vertex induces a knock-on effect for adjacent vertices, such that they also require re-classification. For example, a 4B-vertex is highlighted with an `X' in Figure \ref{fig:interpolate}(b), which changes to a 5-vertex in Figure \ref{fig:interpolate}(c) as we connect a broken bond to a 3-vertex. This correction `propagation' depends on the local environment formed when the broken bonds are created, the subsequent choices made when correcting them, and can be tuned for the desired level of interpolation between structures. Here, we have made choices by hand to keep the changes close to the interface (grey zone). A simple annealing process where vertices are iteratively replaced could be used to build larger scale-interfaces, or to create a more disperse interface zone, for example. 

\begin{figure*}
	\centering
	\includegraphics[width=\linewidth]{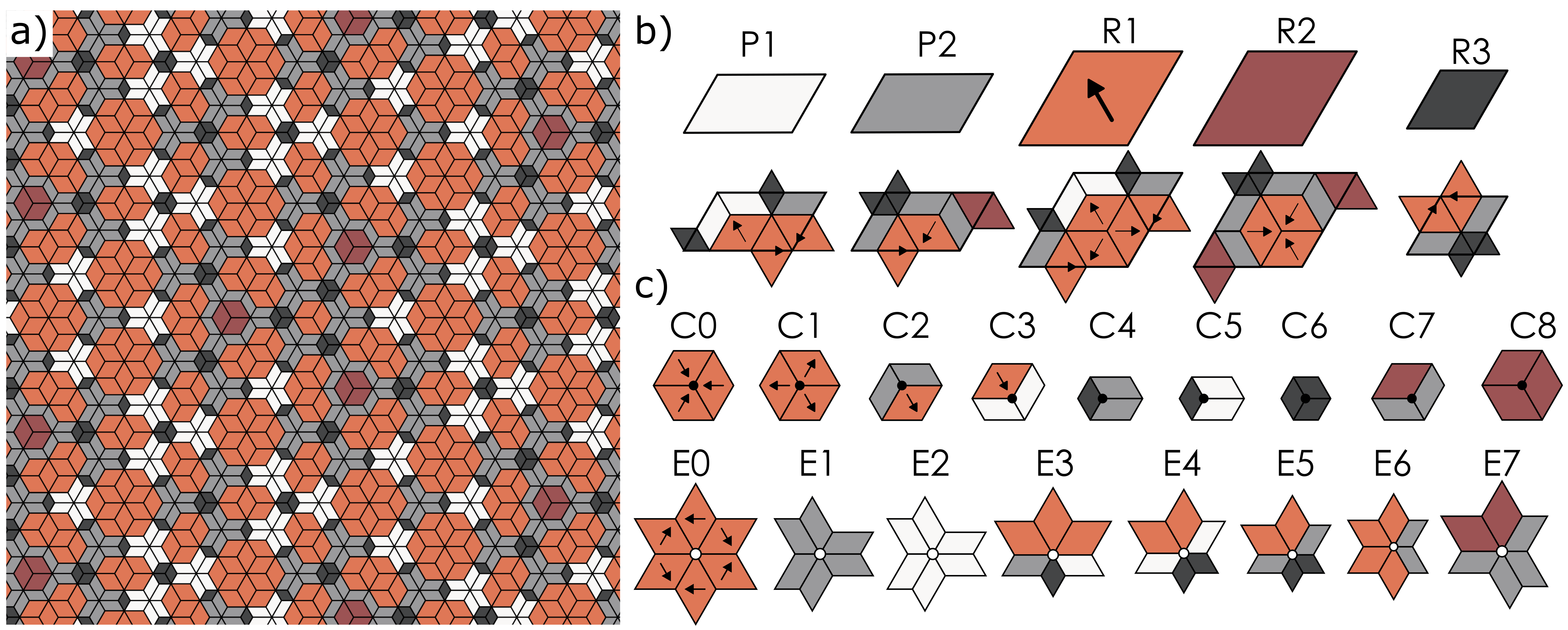}
	\caption{\textbf{(a)} A patch of an aperiodic tiling which is topologically equivalent to the periodic rhombille tiling. \textbf{(b)} Top: the constituent tiles of the aperiodic rhombille tiling, labelled according to their geometry. Bottom: the tile substitution rules, where arrows indicate the relative orientation of the R1 tile. \textbf{(c)} The 17 vertex types of the tiling, where two thirds occupy the black sublattice, and one third occupys the white sublattice. \label{fig:arhombille}}
\end{figure*}

The changes we have made affect the vertex frequencies in either segment: analytical characterisation of local environments at the interface and their effect on resultant properties or models is a topic for further research, beyond the scope of this work. However, as an initial effort along these lines, we show that interfaces can be built from a point which maximises (or minimises) the number of overlapping bonds. Within each $SEH$ tiling there are finite regions of bond distributions which overlap directly with patches of their periodic analogues. To explore the matching behaviour, we generated circular regions of tiling vertex maps: the radii of periodic regions was set to 50, the aperiodic to 25, which comprises roughly 3500 and 1500 vertices respectively. Then, we superimposed the aperiodic maps directly on top of their periodic analogues, and calculated how many vertex bonds overlap. Note that a `full' match is not required i.e. a 3-vertex can sit atop a 6-vertex, where 3 of the bonds directly overlap and are thus counted as 50$\%$ occupancy. Subsequently, the aperiodic discs were shifted so as the centre of the disc visited every site on the underlying triangular lattice. Figures \ref{fig:interpolate}(g, h) show maximal overlap patches for the \SEHzz{} and \SEHhh{} tilings, where $\sim$61$\%$ and $\sim$57$\%$ of bonds match. Figure \ref{fig:interpolate}(i) shows the percentage of overlaps as the discs are moved: the \SEHzz{} has a lower minimal matching, but a higher maximum. The effect of disc/patch size or rotations between the aperiodic/periodic sets with respect to this overlap behaviour are also parameters to be explored further.

\subsection*{Vertex rotations and an aperiodic rhombille tiling}

We can also create an interface through a simple relationship identified between the \SEHhh{} the rhombille tiling. The rhombille tiling consists of 3- and 6-vertices, where every 6-vertex is connected to a 3-vertex, and vice versa. First, we wish to demonstrate the effect of selecting two adjacent 3-vertex nodes in the rhombille tiling (which do not share tiles), and rotating them each by 180$^\circ$. Figure \ref{fig:rhombille}(a) shows how this operation creates a new set of vertices: one 4A-, six 4B-, and four 5-vertices, each highlighted by black circles.

By rotating a specific subset of 3-vertices across the \SEHhh{}, we can transition to a periodic rhombille tiling: this operation can be considered as a collective phason flip which acts as a phase transition between an aperiodic and periodic structure. We discuss the origin of the subset we choose and the relationship to phasons from a superspace perspective in Appendix \ref{app:arhombille}. Figure \ref{fig:interpolate}(b) shows the \SEHhh{} tiling with these vertices indicated by black circles. Figure \ref{fig:rhombille}(c) shows the result of rotating these nodes by 180$^\circ$, where we have retained the original colour scheme of our tiles as a reference. Similarly, see Supplementary File 1 for an animation showing the transition, where we change the colour scheme to accentuate the structural differences. In Figure \ref{fig:interpolate}(c), the connections between vertices are identical to a periodic rhombille tiling, which can be seen by eye -- only the colour scheme remains aperiodic. 

Considering this relationship, interfaces can also be created by rotating groups of vertices belonging to this subset. As an example, Figure \ref{fig:interpolate}(d) shows an \SEHhh tiling where the 3-vertices indicated in Figure \ref{fig:rhombille}(b) within a radius of 10 around the origin have been rotated by 180$^\circ$. Each of the tiles within this radius have been coloured the same for clarity. Again, a seamless interface is created around the perimeter of this disc. Whether a similar method can be applied to the \SEHzz{} tiling is an open question.

As an interesting, unintended consequence, the rotated tile system in Figure \ref{fig:interpolate}(c) can be used to create a new aperiodic tiling which is topologically equivalent to the periodic rhombille tiling. Here, we re-scale the appropriate tile vectors back to their $\tau$ length scales (alternatively, we can perform the rotation operation of our 3-vertex subset on the original \Hhh{} tiling). In either case, we find the tiling shown in Figure \ref{fig:arhombille}(a), where we have changed the colour scheme to reflect the different tiles and their substitution rules, which we show in \ref{fig:arhombille}(b). The tiles are labelled according to their geometry - parallelogram or rhombus, and R1 is marked with an arrow to indicate its relative orientation under substitution. The 17 distinct vertex types of the tiling are shown and labelled in Figure \ref{fig:arhombille}(c), which occupy two sub-lattices indicated by the black (3-vertex) and white (6-vertex) circles. The ratio of 3-vertices to 6-vertices is exactly 2:1, as expected from the rhombille tiling. Proof of vertex frequencies and other tiling properties are detailed in Appendix \ref{app:arhombille}. The aperiodic rhombille tiling also presents an opportunity for creating an interface, where the edge lengths of the relevant tiles are gradually scaled from 1 to $\tau$ in a region of interest, for instance.

%Similarly, and importantly

%We note that the combinatorial sub-lattice selection method we have used here could also be applied to the Fibonacci square tiling, or to the marked supertiles of the `Spectre' monotile, for instance \cite{Lifshitz2002square, Smith2023chiral}.

\section{Flexibility and applications}

The tilings and interfaces we have discussed serve as examples for a broad range of systems that are inherently flexible in their design. Here, we briefly discuss the different options available for either ad hoc or rational design. Then, we move on to discuss a few areas of application immediately relevant to the systems we have introduced. 
\begin{figure}
	\centering
	\includegraphics[width=\linewidth]{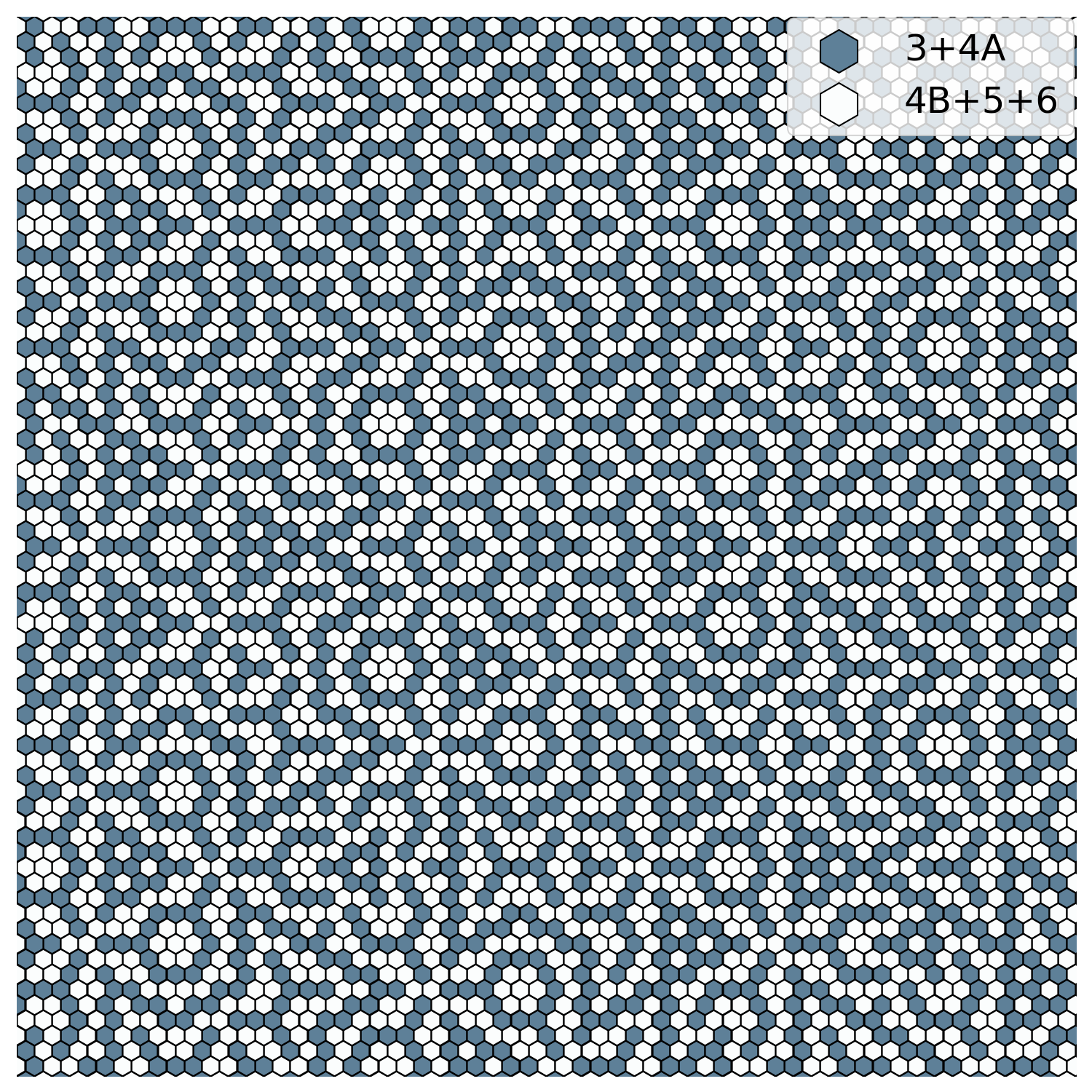}
	\caption{An aperiodic bipartite decoration of a periodic triangular lattice, using the $n$-vertices of the \Hhh{} tiling. The 3-, and 4A-vertices are grouped and plotted together as blue hexagons, with the remaining vertices as white. \label{fig:bipart}}
\end{figure}
\subsection*{Design flexibility}

There are two takeaways in terms of design flexibility: first, both the frequency and arrangement of $n$-vertex maps can be altered by changing the scaling between $\kv^{(1-3)}$ and $\kv^{(4-6)}$. As we have discussed, as long as this scaling is irrational, our distribution of bonds will be aperiodic. Therefore, there is scope to use \textit{any} irrational constant to design aperiodic structures or interfaces either spontaneously, or, with desired structural environments/properties in mind. For instance, larger irrational numbers leads to more frequent intersections between the $\kv^{(1-3)}$ grid family only, which leads to larger periodic-like domains \cite{Matsubara2024aperiodic}. Conversely, smaller numbers lead to more frequent crossings within the $\kv^{(4-6)}$ grid family. Likewise, we have only discussed two examples in terms of grid shift parameters $\alpha_s$ and $\alpha_l$ -- altering these also changes the intersection points in grid space, leading to different geometries and vertex configurations \cite{Coates2024hexagonal}. 
 
Second, the vertex maps of Figures \ref{fig:skele}(c,d) represent a separate starting point for creating a wide range of decorations which can act as stand-alone systems to investigate, with some examples discussed further on. As an example and in the simplest sense, we can construct bipartite systems using distinct sublattices by combining certain $n$-vertex types. For the \SEHzz{} tiling, we have 6 types of vertices that can be combined to give 31 unique two-part structures -- calculated by summing the binomial coefficients and removing the combination where all vertices are selected. Correspondingly, the \SEHhh{} tiling has 15 unique types, one of which is shown as an example in Figure \ref{fig:bipart}. Changing the respective scale of our grid vectors changes these distributions, and more complex decorations can be found by creating tripartite systems etc.

% We note that if we were to consider all instances of the 3-vertices as occupying separate sublattices, the number of available combinations increases dramatically.

\subsection*{Applications}
%
%Here we comment on and discuss a few potential research directions where our tilings and interfaces could be relevant.

\subsubsection{Magnetism}

Artificial spin ices (ASIs) are nanomagnets which decorate lattice geometries, giving rise to often interesting magnetic phenomena \cite{Skjaervo2020advances}. Alongside temperature and disorder, geometry is an important factor in the phase diagrams of ASIs. Indeed, frustrated vertex configurations often give rise to exotic behaviour, where the coordination number of the vertex is an important factor in determining degeneracy, ground, or excited states \cite{Morrison2013unhappy, Gilbert2014emergent}. Similarly, topological defects can play a significant role in any emergent properties \cite{Drisko2017topological}. Both the $SEH$ tilings and the interface systems we have introduced offer high potential for designing and exploring novel ordered ASIs with tunable vertex configurations, coordinations, and topological defects. In fact, recent work explored the effect of disorder in a rhombic ASI, which utilised the exact vertex configurations we discussed in Figure \ref{fig:skele}(c) in a random fashion \cite{Cote2023direct}. Likewise, while several aperiodic ASIs have been constructed, the focus has been on Penrose tilings \cite{Farmer2016direct, Shi2018frustration}, such that research along this direction offers a different angle to explore while keeping some familiarity with periodic systems.

%\cite{Nisoli2013colloquium, Skjaervo2020advances} 

Similarly, there are many avenues for exploring the theoretical magnetic models \textit{on} our systems e.g., as Ising, \textit{xy}, or \textit{xyz} spin systems, or, as to whether there are exact dimer model solutions as with other trigonal/hexagonal and aperiodic tilings \cite{Yan2022triangular, Moessner2001phase, Singh2023hamiltonian, Singh2023exact}. A particularly relevant previous study showed that random fluctuations of vertices in a dice lattice enhanced its antiferromagnetic behaviour \cite{Jagannathan2012geometric}. It would be intriguing to see if this property changes considering our ordered, but aperiodic fluctuations.

On this point, as mentioned in Section \ref{sec:SEH}, the vertex frequencies of our $SEH$ tilings are scale independent, which leads us to expect identical magnetic behaviour to our previous work on the \Hzz{} and \Hhh{} tilings under the Hubbard model \cite{Koga22,Matsubara2024ferromagnetically}. Here, the magnetic states, ferri- and ferromagnetic respectively, were calculated considering equal hopping interactions between vertices separated by non-equal edge lengths. However, the magnetic states are dependent on the vertex frequency imbalance -- which remain identical properties for the $SEH$ tilings, such that scaling our edges to unity does not affect our hopping interactions. 

%

%As a general comment on our theoretical magnetic investigation with a view to future work, we note we have only explored two extremes of a simple toy system. However, the behaviour of the \SEHzz{} tiling at these two extremes (similar to a random defect model at $J_2=-1$, less frustrated than a triangular lattice at $J_2 = 1$) indicates a potentially rich magnetic phase diagram to explore, particularly with ordered/disordered vacancy defects in mind. 

\subsubsection{Metamaterials}

Metamaterials are artificial materials which are engineered to exhibit specific or unusual effective properties, and represent an exciting research field. Fundamentally, metamaterials are composed of some building block (or `meta-atom') which decorates a lattice structure, in either 2D or 3D \cite{Askari2020additive}. The scale/configuration of the lattice and the design/constituent of the building block are parameters that can be varied to yield novel properties to be exploited across a wide range of disciplines. The application of our systems and interfaces could be similarly varied if we consider existing work that explores triangular/hexagonal structures across these disciplines. For example, it is easy to imagine the bipartite decoration in Figure \ref{fig:bipart} representing scatterers in waveguides \cite{Chaunsali2018experimental,Tallarico2017tilted, Wolff2014formal,Zhao2021plane,Perez2019scalable, Cho2012surface, Wang2020valley,Wang2022extended} or as photonic materials with different dielectric constants \cite{Wu2015scheme,Kim2002two, Kim2001strain, Li2018toward}.

Perhaps more directly relevant are mechanical metamaterials \cite{Sinha2023programmable, Mukhopadhyay2017effective, Mukhopadhyay2017stochastic, Valipour2022metamaterials}. Here, structures based on or including triangular or hexagonal features are probably the most widely explored. From a general starting point, these geometries are low density, lightweight, and exhibit excellent or novel mechanical properties as a direct result of their structure e.g. high strength due to their point isotropy or high node connectivity \cite{Yin2023review, Gu2018experimental, Rafsanjani2016bistable,Goldsberry2018negative,Mehreganian2021structural,Grima2012three}. Similarly, trigonal and hexagonal structures have recently served as the basis of designs where the thickness of beams is varied either periodically or randomly in attempts to maximise damage tolerance under loads \cite{Joedicke2024addressing,Ryvkin2020fault,Cherkaev2019damage, Ryvkin2021analysis}. Our designs have potential in this area as the exploration of mechanical aperiodic structures is showing promise \cite{Moat2022compressive, Imediegwu2023mechanical}, yet here we can also retain some of the advantageous aspects of periodicity.

On a related note, we have recently investigated the effects of linear regime water wave propagation through periodic and aperiodic arrays of cylinders, and linked these findings to optical (and acoustic) wave manipulation through a transformation optics approach \cite{Smerdon2024interaction}. In this work, the array based on \SEHzz{} vertices was the most effective at blocking over all explored frequencies, indicating potential for application across other wave-based disciplines.

\subsubsection{Scanning Tunneling Microscopy}

Scanning Tunneling Microscopy can be used to manipulate individual atoms or molecules at a surface, with various applications including studying chemical reactions \cite{Hla2003stm}, creating novel quantum arenas \cite{Crommie1995quantum}, and designing fractional dimension electronic states \cite{Kempkes2018design}. Hexagonal close packed surfaces represent an ideal playground for these types of investigations \cite{Hla2005scanning,Lagoute2004manipulation,Kanisawa2019quantum,Wong2015characterization}, and also represent a basic scaffold for recreating designs such as Figure \ref{fig:bipart}. A clean surface could be decorated with adsorbates according to the pattern formed by the white hexagons, or vice versa. From here, the effect of an aperiodic-periodic interface on the electronic states could be explored, perhaps providing a toy model for further insight into the electronic states in inter-metallic quasicrystals.
%\subsubsection{Diffusion/Percolation}
%\subsubsection{Cellular automata}

%\subsubsection{Soft matter}
%Out of interest, we note that the vertices in Figure \ref{fig:skele}(c) look similar to the schematic representations of patchy particles studied in self-assembly systems \cite{Zhang2004self}. Whether the vertices we present here could theoretically or experimentally self-assemble -- similar to other patchy particle/tiling work \cite{ Whitelam2015emergent, Karner2019design,Karner2020matter,Millan2014self,Millan2015effect} -- is an intriguing question for future work. 

\section{Conclusions}

We have presented a series of methods in which to build coherent periodic to aperiodic interfaces. We produced variations of the $H$ tilings \cite{Coates2024hexagonal}, which produce single edge length aperiodic distributions of rhombic tiles with trigonal and hexagonal symmetry. From here, we analysed the vertex distributions of the $SEH$ tilings, demonstrating that they can be described as aperiodic subsets of a periodic triangular lattice. Then, we demonstrated how interfaces can be created - either by directly sandwiching segments together or by rotating a subset of 3-vertices. As a consequence, we defined a new aperiodic rhombille tiling. 
 
The aperiodic structures and decorations we have presented allow for the investigation of interfacial aperiodic/periodic arrangements which have minimized spatial frustration. These decorations suggest wide and flexible experimental opportunities which could be realised and explored at multiple length scales.

\section{Acknowledgements}

This work was supported by EPSRC grant EP/X011984/1. SC would like to thank J. A. Smerdon and P. Moriarty for useful discussions. 

\bibliographystyle{abbrv}
\bibliography{references}

\setcounter{figure}{0}
\renewcommand{\thefigure}{A\arabic{figure}}

\appendix
\clearpage
\onecolumngrid
%
%
%\section{Vertices}
%
\section{\textit{SEH} tiling duals} \label{app:deletion}

The inset of Figure \ref{fig:skele}(b) shows that an alternative way to produce the \SEHhh{} tiling would be to remove a subset of edges from a periodic triangular lattice. The centre point of these removed edges sit at the centre of the \SEHhh{} tiles, such that if we produce the dual of the \SEHhh{} tiling, its vertices consist of `removal' points: overlaying this dual on top of a periodic triangular tiling provides a guide for which edges to remove in order to produce the \SEHhh{} tiling. The \SEHzz{} vertices also decorate a triangular lattice, albeit with additional aperiodically spaced `vacancies' which arise from the centre of the hexagonal tiles. These vacancies are aperiodically ordered defects in a triangular lattice, and, similar to the \SEHhh{} tiling, we can produce a dual using the centres of the removed edges.

\begin{figure}[h!]
	\centering
	\includegraphics[width=\linewidth]{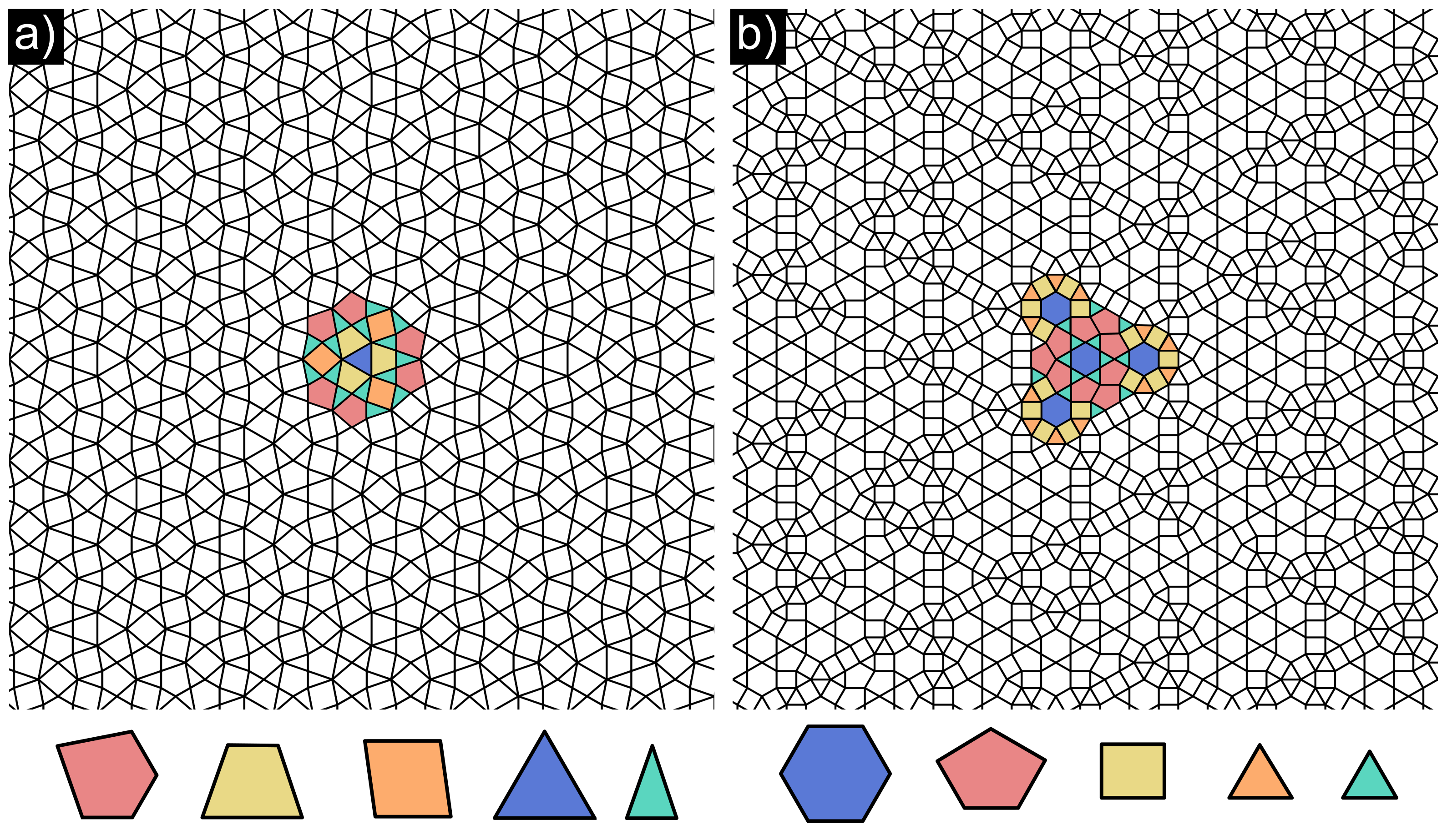}
	\caption{Top: duals of the \SEHzz{} and \SEHhh{} tilings (\textbf{(a, b)} respectively), where the vertices of the duals correspond to the centre of the $SEH$ tiles. Selected tiles are coloured to emphasise the tiling constituents. Bottom: enlarged versions of the dual tiles. \label{fig:duals}}
\end{figure}

Figure \ref{fig:duals} shows the dual of the $SEH$ tilings, or, the edge removal guide to be overlaid onto a periodic triangular lattice - as discussed in the main text. The constituent tiles are coloured for the \SEHzz{} and \SEHhh{} duals  ((a, b) respectively), and enlarged at the bottom of the figure.

\clearpage
\section{Vertex rotation and an aperiodic rhombille tiling \label{app:arhombille}}
\subsection{3-vertex selection}
\begin{figure}[h!]
	\includegraphics[width=\linewidth]{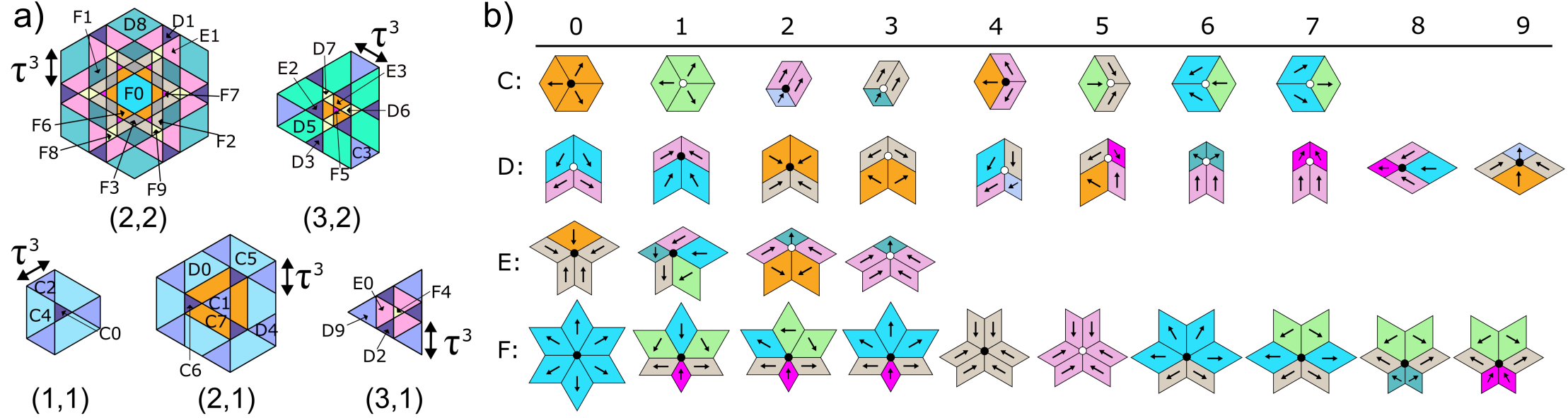}
	\caption{\textbf{(a)} Five of the nine perpendicular-space windows which correspond to the real-space vertices of the \Hhh{} tiling. The windows are labelled as ($h_1$, $h_2$) to indicate the heights at which they intersect two 3D cubes which occupy a 6D superspace. Each window is subdivided and colour-coded to indicate which type of vertex configuration they represent. Arrows and $\tau$ mark the scale of the windows and their sub-domains. \textbf{(b)} The 32 vertex types of the \Hhh{}, labelled according to their coordination number (C, D, E, F). Reproduced and modified from \cite{Coates2024hexagonal}, which describes the superspace analysis and vertex configurations in greater detail. \label{fig:hhsuperspace}}
\end{figure}
\noindent To transition from the \SEHhh{} tiling to the periodic rhombille tiling (or vice versa), we need to rotate a set of 3-vertices, according to Figure \ref{fig:rhombille}(a). We first find all 3-vertices by considering the superspace approach for the \Hhh{} tiling in \cite{Coates2024hexagonal}. As mentioned in the main text, we can still analyse our $SEH$ vertices using this approach, as the 6-dimensional indices which correspond to each vertex remains identical. Figure \ref{fig:hhsuperspace}, reproduced from \cite{Coates2024hexagonal}, shows 5 of our 9 windows, where the remaining 4 are mirrored about the central window. Each window is labelled with coordinates, ($h_1$, $h_2$), corresponding to the `heights' of planes which intersect two 3D cubes occupying a 6D superspace -- see [section V, \cite{Coates2024hexagonal}] for more details. Here, all of our vertex types occupy specific sub-domains on our perpendicular space windows: the eight vertices which have a coordination number of 3 are labelled as `Cx', where x is an integer. These vertices are shown in Figure \ref{fig:hhsuperspace}(b).

The necessary vertices to transform (those decorated in Figure \ref{fig:rhombille}(b)) uniquely occupy the (1,1) and (3,3) windows, or correspond to the C0; C2; C4 vertex types in the \Hhh{} tiling. These were found by inspecting the local environment of all 3-vertex types i.e., by considering which vertex types they are connected to. By intuition, we required that each 3-vertex on the \Hhh{} tiling which was already connected directly to a 6-vertex should remain unchanged, such that we maximise the topological resemblance to a periodic rhombille tiling (C1, C3, C5, C6, C7). Consequently, we found that the remaining 3-vertices (C0, C2, C4) occupy the (1,1) and (3,3) windows.

\subsection{Aperiodic rhombille tiling properties}

\noindent By rotating the C0, C2, C4 vertices of the \Hhh{} tiling 180$^\circ$, or by re-scaling the appropriate tile edges of the \SEHhh{} after the same operation, we obtain the aperiodic rhombille tiling as in Figure \ref{fig:arhombille}. Considering the substitution rules in Figure \ref{fig:arhombille}(b), We can obtain the tile frequencies using the substitution matrix \cite{Senechal96}:

\begin{equation}\label{Eq:M00}
	M = 
	\kbordermatrix{
		& P1 & P2 & R1 & R2 & R3 \\[0.2em]
		P1 & 2  & 0 & 2 & 0 & 0 \\[0.2em]
		P2 & 1  & 3 & 2 & 4 & 2 \\[0.2em]
		R1 & 2  & \frac{3}{2} & 4 & 3 & 1 \\[0.2em]
		R2 & 0  & \frac{1}{2} & 0 & 1 & 0 \\[0.2em]
		R3 & 1  & 1 & 1 & 1 & 1 \\
	},
\end{equation}

\noindent whose largest eigenvalue is $\tau^4$, corresponding to the eigenvector $(1, \sqrt{5}, \frac{3 + 3 \sqrt{5}}{4}, \frac{3 - 3 \sqrt{5}}{4}, 1)$, such that for every $P1$ tile, there are $\sqrt{5}$ as many $P2$ tiles, etc.

We can use a superspace approach to calculate the vertex frequencies of the 17 configurations shown in Figure \ref{fig:arhombille}(c). To do so, we start with a brief description of the vertices of the \Hhh{} tiling. Each vertex is determined by $\sum_{j}^{}n_j\av^{(j)}$, where $n_j$ is a 6-tuple of integers corresponding to the signed, indexed spaces between grid lines in the $j$th grid, and $\av^{(j)}$ are the tiling vectors \cite{deBruijn86,Coates2024hexagonal}. Then, to find the ($h_1$, $h_2$) coordinates of a particular vertex, following Socolar \cite{Socolar89}: $h_1 = e_1\cdot n$, $h_2 = e_2\cdot n$, where $e_1 = [1,1,1,0,0,0]$ and $e_2 = [0,0,0,1,1,1]$ are basis vectors of our two 3D cubes, such that we are projecting each point onto the body-diagonals of these cubes [section IV, \cite{Coates2024hexagonal}].

\begin{figure}
	\centering
	\includegraphics[width=\linewidth]{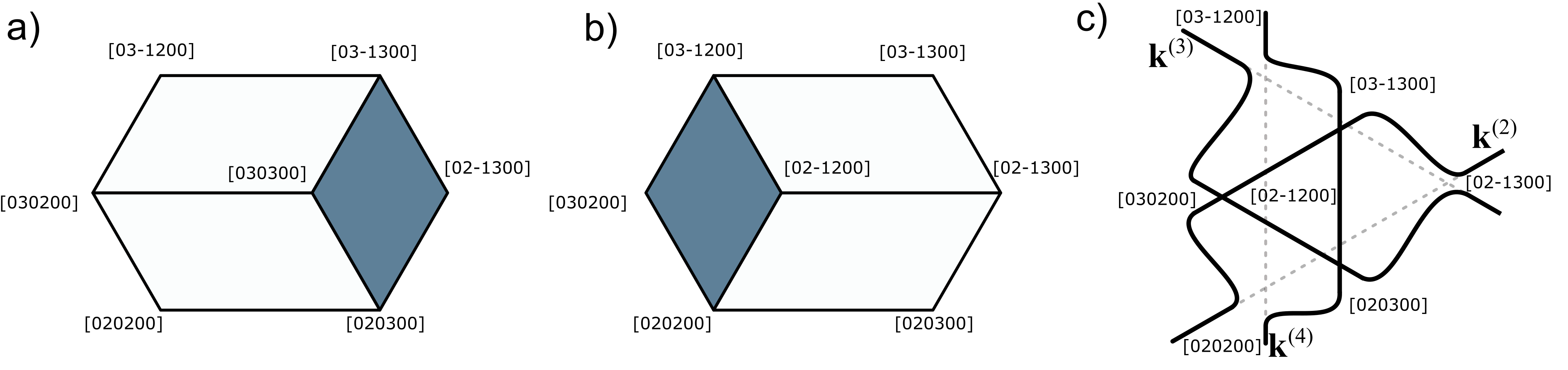}
	\caption{\textbf{(a)} A 3-vertex from the \Hhh{} tiling, decorated with indices which correspond to the tile vertices in 6D superspace. \textbf{(b)} The same 3-vertex after rotation, or phason flip, with new index values. \textbf{(c)} The dual-grid view of \textbf{(b)}, which shows how the grids can be deformed locally to produce our flipped vertices. Grey dashed lines indicate the original grids.\label{fig:flip}}
\end{figure}

Figure \ref{fig:flip}(a) shows a 3-vertex from the \Hhh{} tiling, or a vertex type C2, generated directly using the dual-grid method. The vertices of each tile are labelled with their corresponding $n$ integers, such that we can travel from the [020200] point to each vertex using the tiling vectors \av$^{j}$. It is simple to calculate that the 3-vertex position will reside in the (3,3) window following the method described above. Figure \ref{fig:flip}(b) shows the effect of rotating the 3-vertex 180$^\circ$, which is analogous to a phason flip in a tiling or quasicrystal \cite{Henley1999random,Henley2006discussion}. After flipping, the 3-vertex is necessarily described using a different set of indices, as the tiling vectors needed to move from the [020200] point to the 3-vertex have changed. The grid-space view of this flip is shown in Figure \ref{fig:flip}(c), where the grid lines are in black with their corresponding family labelled as \kv$^{(j)}$, the 6-tuples indicate the indexed spaces between grid lines, and the pre-flip lines are indicated as grey dashes. When the vertex flips, we are effectively locally distorting the grid lines in such a way that the indexed space corresponding to the 3-vertex changes from [030300] to [02-120]. After performing the flip, our ($h_1$, $h_2$) value for Figure \ref{fig:flip}(a) changes to (1, 2).

\begin{figure}
	\centering
	\includegraphics[width=.5\linewidth]{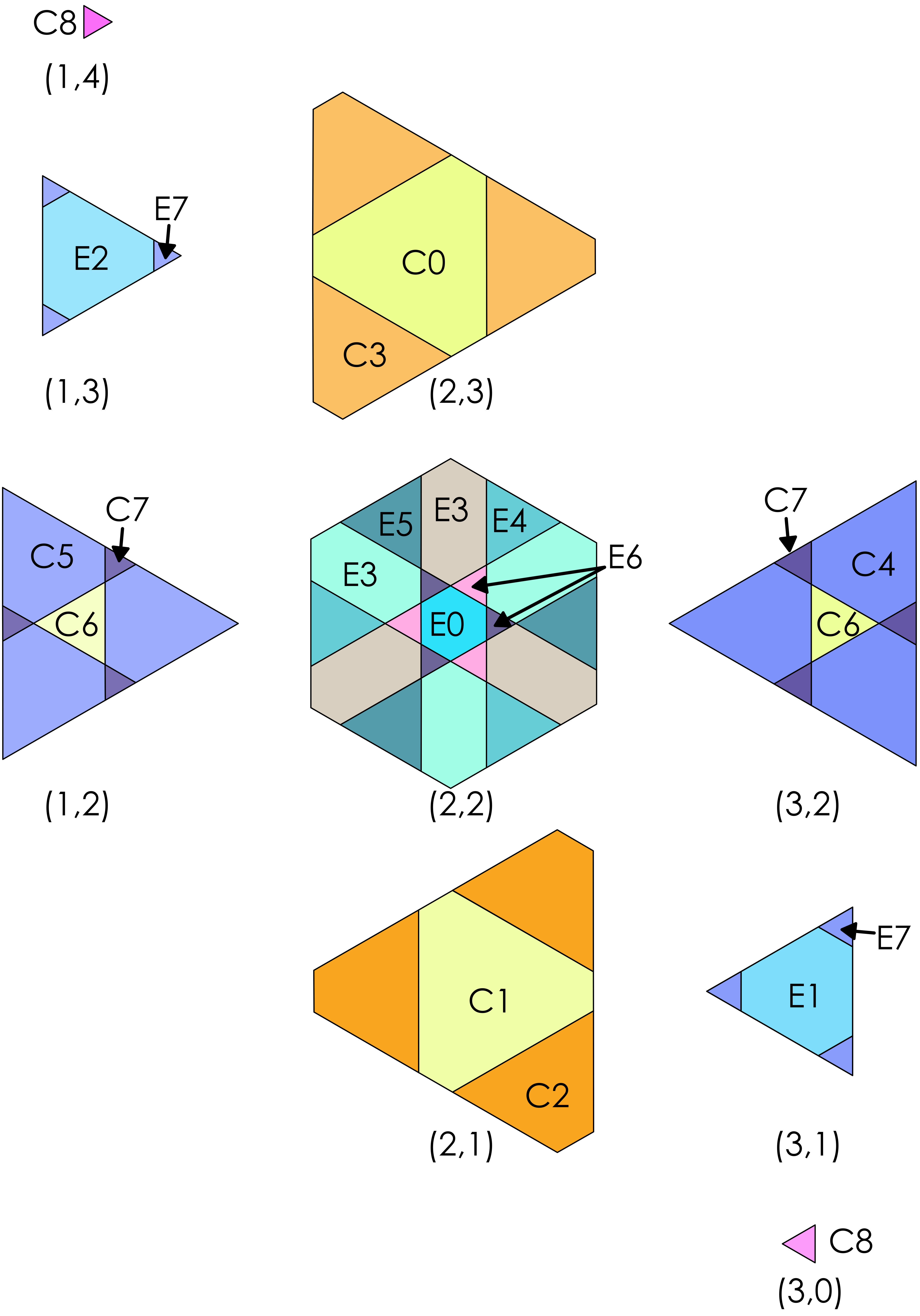}
	\caption{The perpendicular space windows of the aperiodic rhombille tiling. The coordinates of the windows ($h_1, h_2$) are calculated by $h_1 = e_1\cdot n$, $h_2 = e_2\cdot n$, where $n$ are the 6-tuple of indices corresponding to each vertex, and $e_1$ and $e_2$ are the basis vectors of two 3D cubes in 6D superspace. The windows are split into sub-domains representing the specific vertex configurations, which are also labelled. \label{fig:arhombperp}}
\end{figure}

Figure \ref{fig:arhombperp} shows our perpendicular space windows for the aperiodic rhombille tiling, obtained after performing flips for all C0, C2, and C4 vertices on the \Hhh{} tiling. The windows at (1, 1) and (3, 3) disappear, two new windows at (3, 0) and (1, 4) appear, and the shapes of the remaining windows change compared to the \Hhh{} tiling. Setting the smallest window area to 20 to scale our values, we find that our window areas are:
\begin{equation}\label{Eq:windowAreas}
	\textstyle
	\normalsize
	\begin{aligned}	
		(3, 0):&\, 4\sqrt{5}+21 \\
		(2, 1):&\, 590\sqrt{5}+1329 \\
		(3, 1):&\, 120\sqrt{5}+280\\
		(1, 2):&\, 360\sqrt{5}+824 \\
		(2, 2):&\, 726\sqrt{5}+1633 \\
		(3, 2):&\, 380\sqrt{5}+861 \\
		(1, 3):&\, 108\sqrt{5}+261 \\	
		(2, 3):&\, 574\sqrt{5}+1293 \\
		(1, 4):&\, 20 \\
	\end{aligned}
\end{equation}
\noindent Each window can be subdivided to represent the 17 vertex configurations, and we can calculate their frequencies by considering the fractional area of these sub-domains as a sum of the total area (2862$\sqrt{5}$+6522), such that:
\begin{equation}\label{Eq:subAreas}
	\textstyle
	\normalsize
	\begin{aligned}	
		C0:&\, 241\sqrt{5}+549\simeq 8.4\% \quad &	E0:&36\sqrt{5}+91\simeq 1.3\%\\
		C1:&\, 233\sqrt{5}+1329\simeq 8.1\%\quad & E1:&108\sqrt{5}+217\simeq 3.6\%\\
		C2:&\, 333\sqrt{5}+280\simeq 11.5\%\quad & E2:&\frac{1}{2}(189\sqrt{5}+441)\simeq 3.3\%\\
		C3:&\, 357\sqrt{5}+824\simeq 12.4\%\quad & E3:&444\sqrt{5}+966\simeq 15.2\%\\
		C4:&\, \frac{1}{2}(633\sqrt{5}+1449)\simeq11.3\%\quad & E4:&\frac{1}{2}(195\sqrt{5}+441)\simeq 3.4\%\\
		C5:&\, \frac{1}{2}(620\sqrt{5}+1401)\simeq10.8\%\quad & E5:&\frac{1}{2}(225\sqrt{5}+525)\simeq 4\%\\
		C6:&\, 56\sqrt{5}+155\simeq2.2\%\quad & E6:&36\sqrt{5}+93\simeq 1.3\%\\	
		C7:&\, 42\sqrt{5}+105\simeq1.5\%\quad & E7:&\frac{1}{2}(51\sqrt{5}+207)\simeq 1.2\%\\
		C8:&\, 4\sqrt{5}+41\simeq0.4\% \quad & \\
	\end{aligned}
\end{equation}
\noindent where the sum of the C and E vertices represent exactly $\frac{2}{3}$ and $\frac{1}{3}$ of all tiling vertices, respectively.

\end{document}